\documentclass{moriond}

\usepackage{amsmath,amssymb} 
\usepackage{siunitx}

\def\Journal#1#2#3#4{{#1} {\bf #2}, #3 (#4)}

\def\lemaitre{\textsc{Lema\^itre}\ }

\def\be{\begin{equation}}
\def\ee{\end{equation}}
\def\bea{\begin{eqnarray}}
\def\eea{\end{eqnarray}}

\newcommand{\Photo}{}

\begin{document}
\vspace*{4cm}
\title{On the importance of Earth's atmosphere for SNIa precision cosmology}

\author{J\'er\'emy Neveu$^{1,2}$ on behalf of the \lemaitre collaboration and AuxTel collaborations}

\address{$^{1}$Sorbonne Universit\'e, CNRS, Universit\'e de Paris, LPNHE, 75252 Paris Cedex 05, France; $^{2}$Universit\'e Paris-Saclay, CNRS, IJCLab, 91405, Orsay, France.}

	
\maketitle\abstracts{
The measurement of colours in photometric surveys is a path to get access to cosmological distances. But for future large surveys like the Large Survey of Space and Time undertaken by the Vera Rubin Observatory in Chile, the large statistical power of the promised catalogues will make the photometric calibration uncertainties dominant in the error budget and will limit our ability to use it for precision cosmology. The knowledge of the in situ atmospheric transmission for each exposure, each season and on average for the entire survey can help reaching the subpercent precision for magnitudes. I will show the impact of precipitable water vapour during a cosmological survey on supernova cosmology. Then I will present how AuxTel at the Vera Rubin Observatory have the capability to measure the on-site atmospheric transmission  in real time to improve LSST photometric calibration for precision cosmology.
}

\section{Type Ia supernova cosmology with LSST}

The Legacy Survey of Space and Time (LSST) undertaken at the Vera Rubin Observatory (Chile) is a photometric multiprobe cosmological survey. Thanks to a wide-field camera of 9.6 square degrees equipped with 3.2 billion pixels, it will provide a wide, fast and deep survey in 6 bands ($u,\ g, \ r,\ i,\ z$ and $y$, from visible to near-infrared). The full austral sky will be scanned every three days on average with a $5\sigma$ magnitude limit of 24.05 in $r$ band for a single \SI{15}{\s} exposure. After ten years, millions of transient events like type Ia supernovae (SNIa) will be harvested. The camera has been delivered to Chile in May 2024, and the commissioning survey is about to start with the Commissioning Camera (ComCam) in July 2024. The LSST main camera will see its first photon in spring 2025 to start the scientific survey at the end of 2025.

In addition to the wide fast deep survey (WFD), 5 Deep Drilling Fields (DDF) will be probed at the level of 7\% of the total survey time. Cadence is not decided yet, but with the current baseline scenario, the Cadence Working Group gets a million SNIa after ten years of survey with good quality cuts, and \SI{35000} may get spectra for typing and spectroscopic redshifts. Among them, 20\;000 are expected in DDFs, with hundreds of SNIa above redshift $0.8$. Therefore LSST has the capability to provide a full supernova sample from low to high redshift, with 10 times more SN than current surveys.

Here, one has to recall that a SNIa Hubble diagram is interpreted using the $D_L(z)$ (the standard luminosity distance) if SNIa fluxes are standardised beforehand. Historically, the standardised magnitude was chosen as the Bessel $B$ band SNIa rest-frame magnitude $B$ at the maximum date, corrected by $B-V$ colour. Doing so, we always consider the same portion of the SNIa Spectral Energy Density (SED) at the same moment of its evolution and whatever the redshift (Fig.~\ref{fig:SN_atm} left). The distance moduli $\mu(z)$ links cosmological models to the measured fluxes via:
\begin{equation}
\mu(z) = 5 \log_{10}\left[ D_L(z)\right] \approx B - 3(B-V) 
\end{equation}
where we simplify the usual standardization formula in order to keep only the chromatic term, the most impacted by the atmospheric transmission variations. Looking a this formula, the price to pay is the necessity to be able to convert observed $ugrizy$ magnitudes into this rest-frame $B$ band magnitude. For example, the rest-frame $B$ band (resp. $V$ band) of a low-redshift supernova corresponds approximately to the $g$ band (resp. $r$ band), while for a redshift 1 SNIa the $i$ and $z$ bands are concerned. Depending on its redshift, a supernova is thus interpreted using different telescope filters. Therefore, the inter-calibration of these filters is crucial to avoid biasing the distance moduli of the high-redshift SNIa sample compared to the low-redshift sample, which finally can bias the cosmological interpretation of such Hubble diagram.

\vspace*{-2.5mm}
\section{Impact of atmospheric transmission seasonal variations on cosmology}

\begin{figure}[h!]
\vspace*{-0.5cm}
\includegraphics[width=0.48\textwidth]{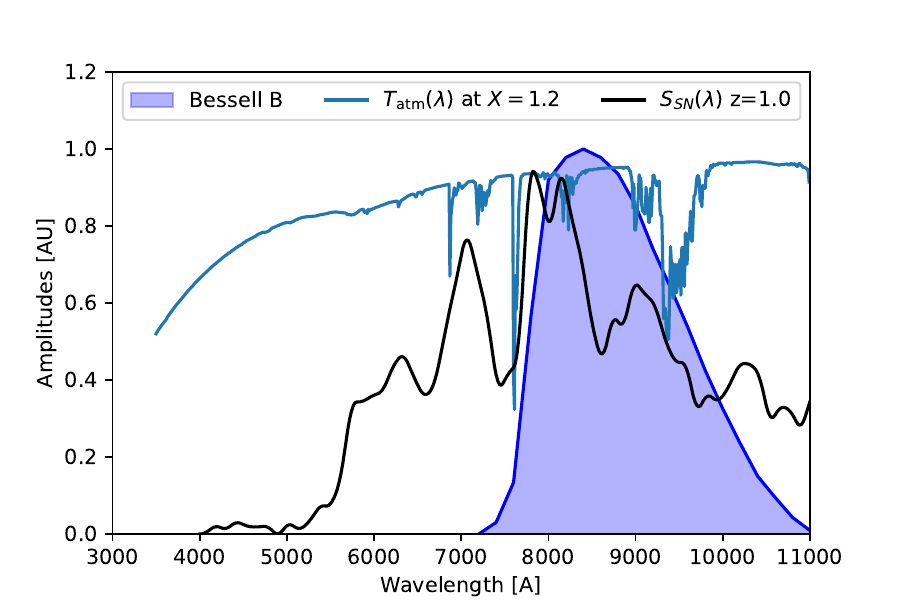}
\includegraphics[width=0.48\textwidth]{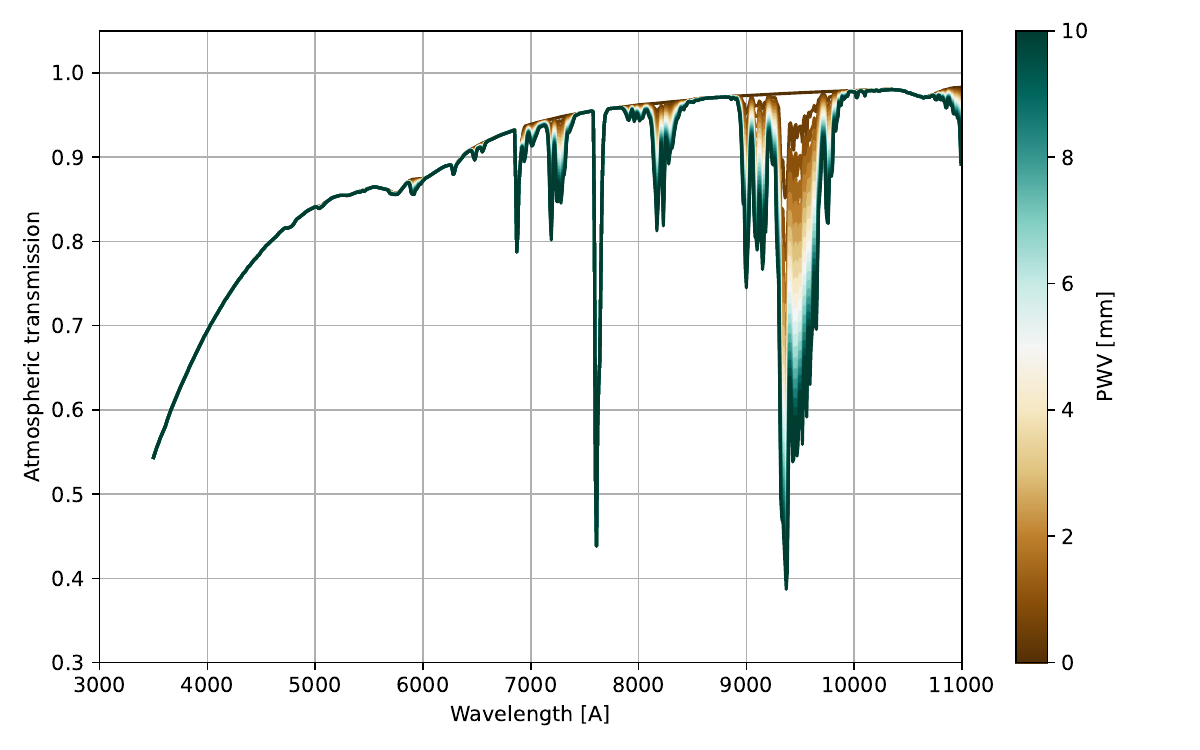}\hfill
\caption{Left: typical atmospheric transmission (blue) at airmass $X=1.2$, type Ia supernova SED at redshift $z=1$ and Bessel B filter transmission redshifted at $z=1$. Right: atmospheric transmission curves coloured by PWV values as a function of wavelength. }\label{fig:SN_atm}
\end{figure}

The effective filter transmission on Earth is the product of the intrinsic filter and telescope throughputs and of the atmospheric transmission. However, all the effective bands are not affected similarly by atmospheric transmission. An inaccurate measurement of the observatory site's mean atmospheric transmission can affect differently low and high redshift supernovae, leading to distortions in the Hubble diagram $\mu(z)$, and bias the cosmological parameters.

Atmospheric transmission as a function of wavelength $\lambda$ is mainly explained by Rayleigh scattering in $\lambda^{-4}$, which only depends on local pressure (Fig.~\ref{fig:SN_atm} right). Ozone absorption is principally marked by a wide absorption band around \SI{600}{\nm} and aerosol presence bends the transmission curve in the ultraviolet part. Precipitable Water Vapor (PWV) creates deep absorption bands around \SI{720}{\nm}, \SI{820}{\nm} and \SI{950}{\nm}. Their deepness depends non-linearly on airmass and on the amount of precipitable water, and can change daily and seasonally.

\begin{figure}[h!]
\includegraphics[width=0.48\textwidth]{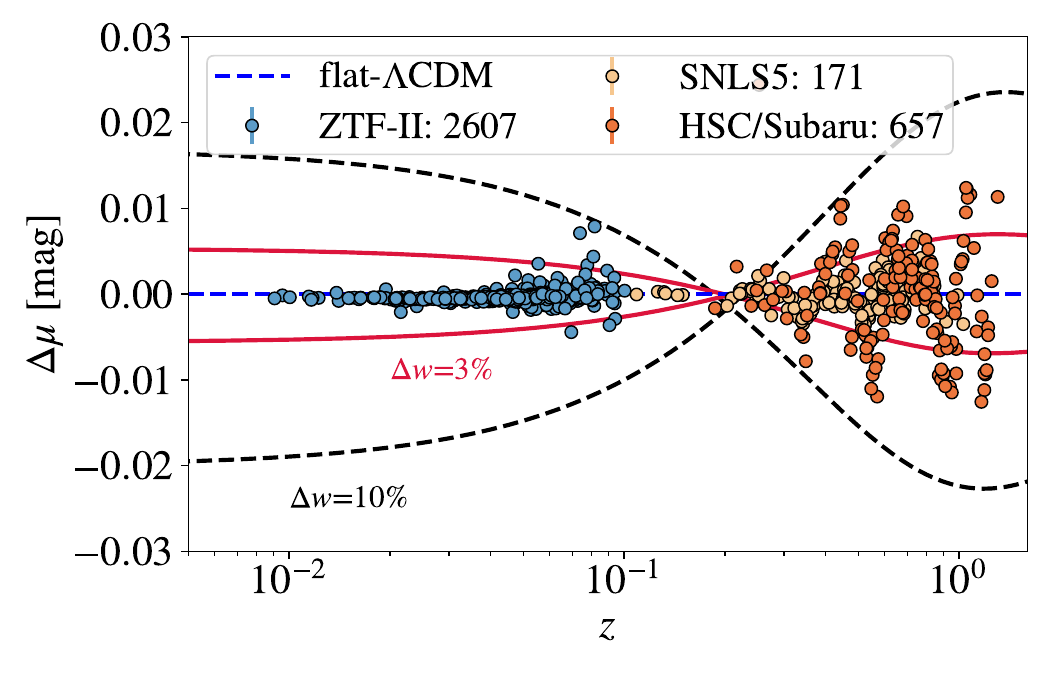}\hfill
\includegraphics[width=0.48\textwidth]{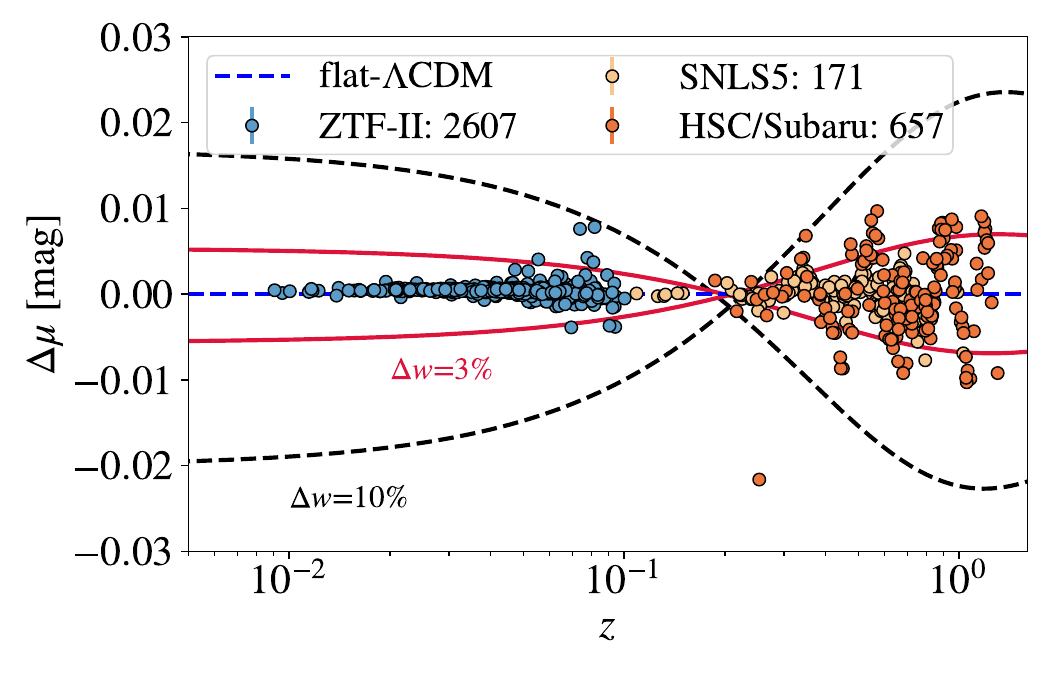}
\caption{Variations $\Delta \mu$ of the distance moduli as a function of redshift when simulated supernovae lightcurves done with $PWV=\SI{5}{\mm}$ are fitted with $PWV=\SI{3}{\mm}$ (left) or $PWV=\SI{7}{\mm}$ (right) for the \textsc{Lemaitre} sample.}\label{fig:deltamu_pwv}
\end{figure}

We simulated the impact of PWV parameter variation on the \textsc{Lemaitre} SNIa sample, which links observations by the ZTF-II, SNLS5 and HSC surveys. Starting from a noise-free simulation of the sample light-curves using $PWV=\SI{5}{\mm}$, we used the SALT2.4 model\cite{salt2} to interpret them with wrong atmospheric transmissions. On Fig.~\ref{fig:deltamu_pwv}, we showed the distortions of the Hubble diagram $\Delta \mu(z)$ when using an atmospheric transmission with $PWV=\SI{3}{\mm}$ (left) and $PWV=\SI{7}{\mm}$ (right). This amplitude was choosen as it may corresponds to seasonal variations (Fig.~\ref{fig:pwv}). High-redshift supernovae are more affected depending on which water absorption band enters the redshifted $B$ band window, leading to a scatter depending on redshift of about 10~mmag. From this toy model, we can conclude that measuring survey mean PWV value better than $\approx\SI{0.2}{\mm}$ can be enough to reach millimagnitude accuracy.

\section{AuxTel, a dedicated slitless spectrograph to monitor atmospheric transmission}

To get atmospheric transmission in real-time, the Vera Rubin Observatory is equipped with an Auxiliary Telescope (AuxTel\cite{auxtel}) dedicated to this mission. Thanks to a slitless spectrograph, it performs spectrophotometric measurements of calibration stars from the CALSPEC catalogue\cite{calspec}. Comparing these out-of-atmospheric spectra with local measurements we can infer the local atmospheric transmission on the line of sight. There isn't a CALSPEC star per LSST field, but, in principle, observing these CALSPEC stars close to the line of sight of the main LSST telescope can provide real-time atmospheric parameters via an atmospheric model like \texttt{Libradtran}\cite{libradtran}.

\begin{figure}[h!]
\centering
\includegraphics[height=5.5cm]{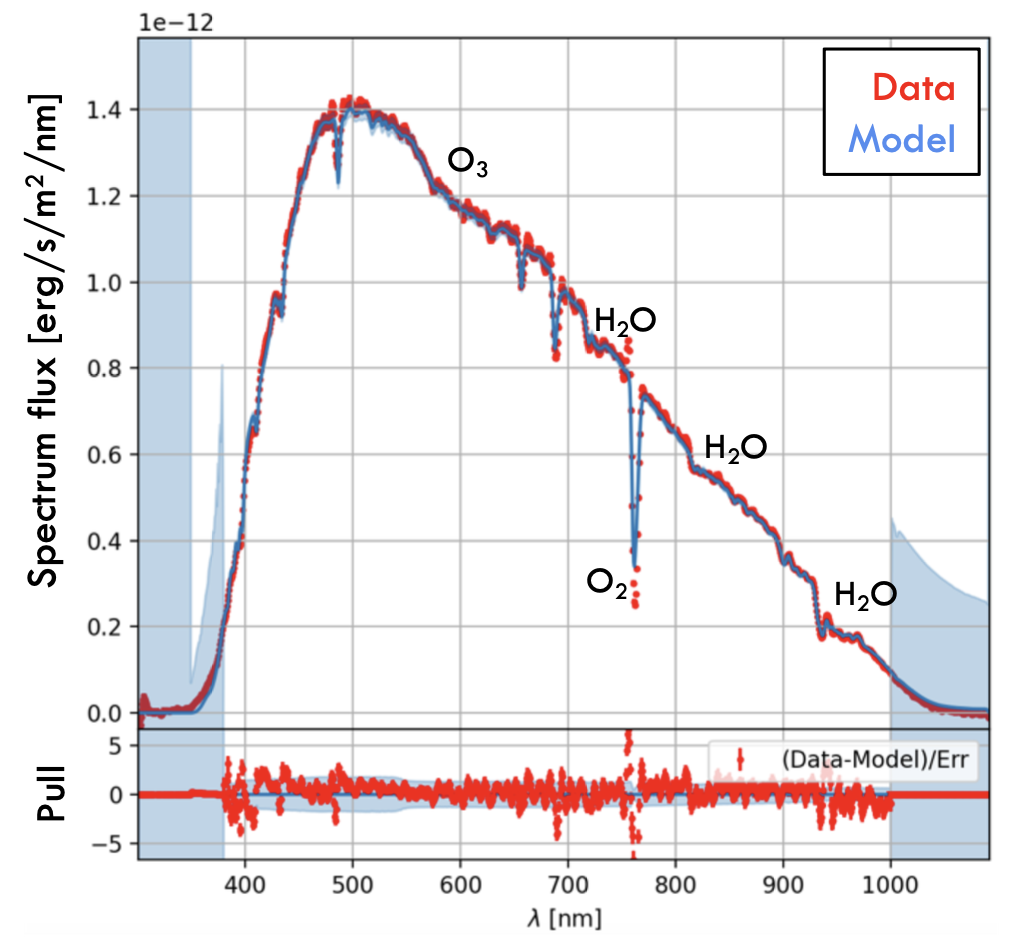}\caption{Reduced Auxtel spectrum (red) of CALSPEC star HD160617 fitted by a model (blue) including CALSPEC SED, AuxTel throughput and a \texttt{Libradtran} model with free atmospheric parameters.}\label{fig:spectrum}
\end{figure}

The AuxTel reduction pipeline is based on \texttt{Spectractor}\cite{spectractor}. Spectrum extraction is performed via forward modelling technique and regularized deconvolution. This approach provides a sufficient accuracy in recovering the exact spectrum on simulations, as soon as the spectrograph is correctly model in terms of dispersion, transmission and Point Spread Function (PSF). An additional benefit of the forward modeling approach is that it provides spectra to test the pipeline. 

\begin{figure}[h!]
\includegraphics[height=7cm]{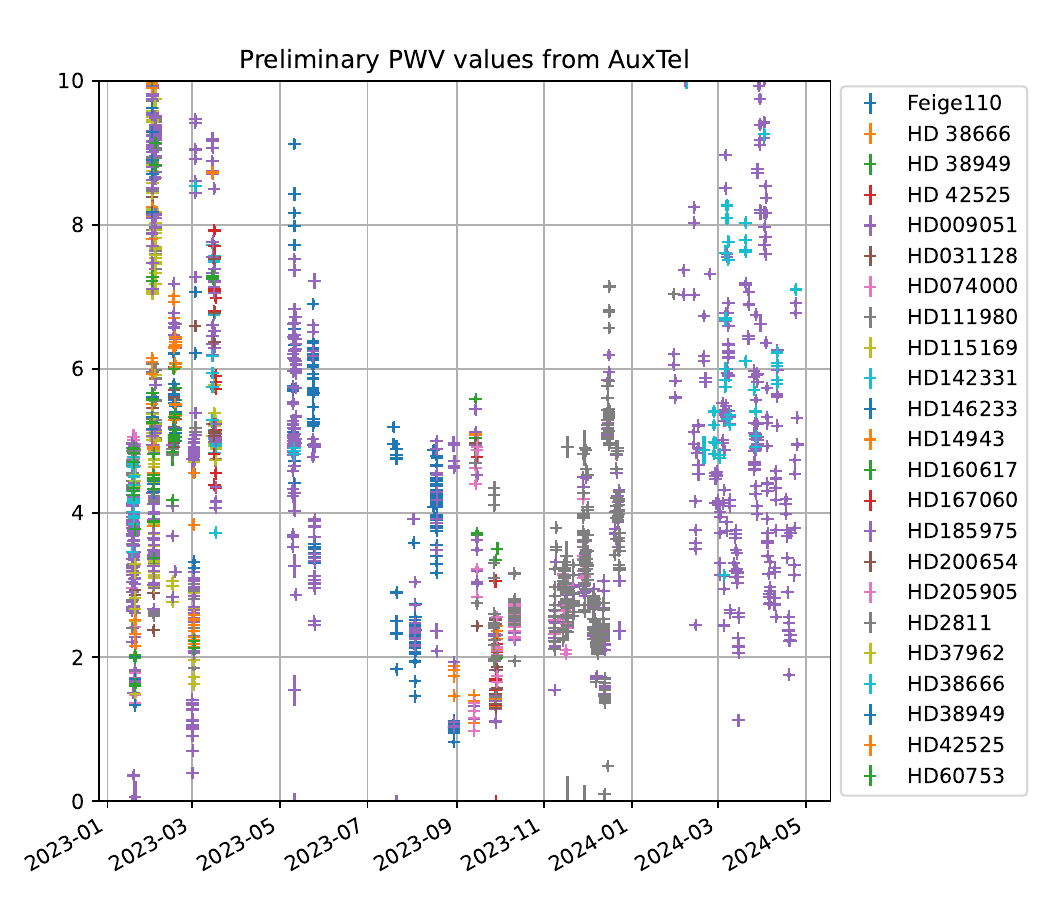}\hfill
\includegraphics[height=7cm]{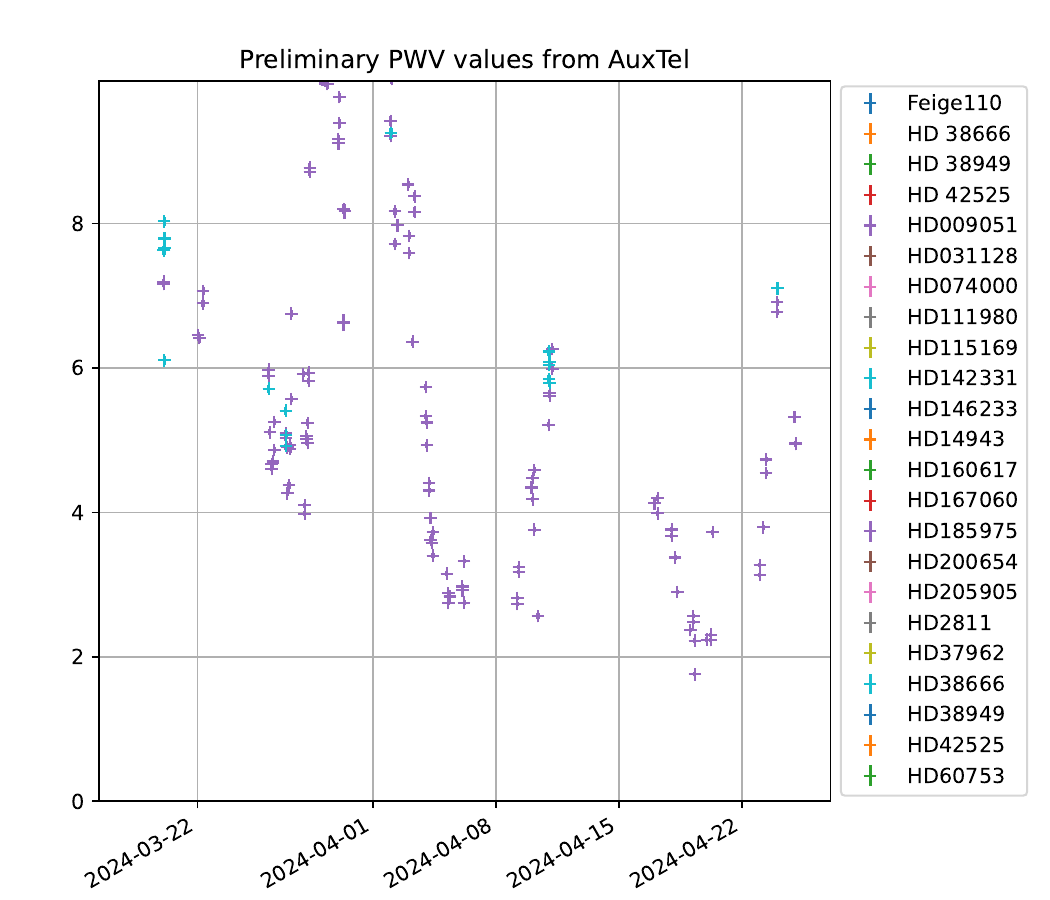}
\caption{Left: PWV fitted parameters on all AuxTel data since 2023/01 as a function of time, coloured by CALSPEC names. Right: zoom on April 2024, showing daily and weekly variations.}\label{fig:pwv}
\end{figure}

The pipeline has been used to extract all AuxTel spectra since January 2023. An example of an extracted calibrated spectrum is represented in red in Fig.~\ref{fig:spectrum}. Absorption bands from dioxygen and water are clearly visible with a spectral resolution sufficient to fit an atmospheric model (blue), multiplied by a prior knowledge of the star SED and of the AuxTel throughput. After 18 months of observation, PWV values exhibit clear seasonal and weekly variations (Fig.~\ref{fig:pwv}). Around October we see a dry and stable season, while in March we have a more unstable season where PWV can change significantly within a week, of the order of what was shown in Fig.~\ref{fig:deltamu_pwv}.

\begin{figure}[h!]
\includegraphics[width=0.48\textwidth]{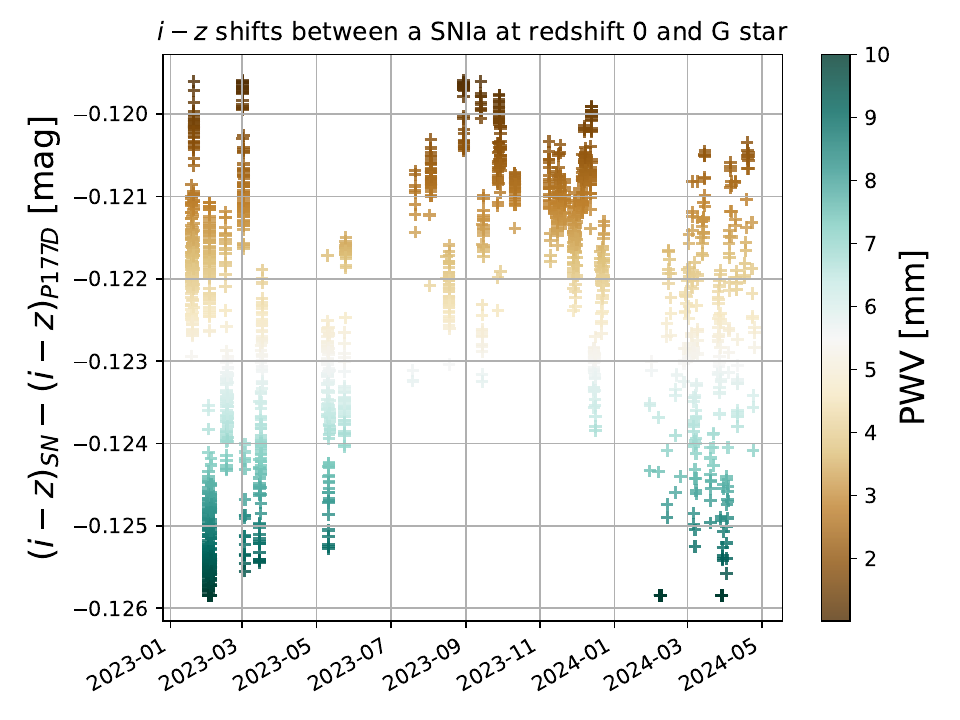}\hfill
\includegraphics[width=0.48\textwidth]{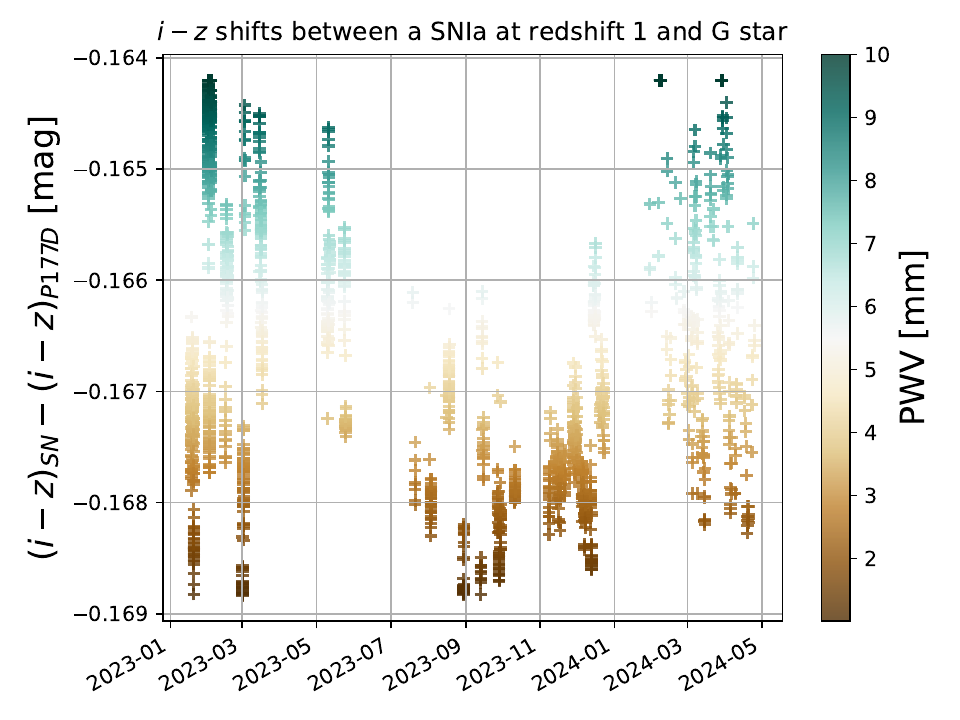}
\caption{$i-z$ colour differences between a type Ia supernova at redshift 0 (left) or 1 (right) and an average G star as a function of time and depending on PWV variations (colour scale).}\label{fig:sn_pwv}
\end{figure}

For supernova cosmology, the measured PWV variations can be translated into SN colour variations compared to field stars (assuming that the survey is calibrated with field star photometry). We used a G star SED as a mean representative for the field stars used to calibrate the SN Ia light curves. For type Ia supernovae at every redshift, the PWV variations measured from 0 to \SI{10}{\mm} induce type Ia colour variations of about 5 millimagnitudes (Fig.~\ref{fig:sn_pwv}). Interestingly, colour variations are inverted for low and high redshift supernovae, meaning that within a season like in October low redshift SNe seem redder while high redshift supernovae seem bluer. Therefore, to correct daily LSST SN photometric measurements to reach millimagnitude accuracy, we showed that it is sufficient to measure PWV at the millimetre level. 

In summary, AuxTel will be used to monitor on-site LSST atmospheric transmission, annually, seasonally and daily. It is necessary to reduce photometric calibration systematics to the millimagnitude level so that supernova cosmological analysis benefit from the full statistical power of the LSST sample. Concerning water absorption, AuxTel performances are accurate enough to reach this goal as shown in Fig.~\ref{fig:pwv}. Further work has to be conducted to reach same accuracy for ozone and aerosol variations.

\end{document}